 \documentclass[final,5p,times,twocolumn,number]{elsarticle}

\usepackage{amssymb}
\usepackage{amsmath}
\usepackage[english]{babel}

\usepackage{graphicx}
\usepackage{dcolumn}
\usepackage{url}
\usepackage{xcolor}
\usepackage{xurl}
\usepackage[
  colorlinks=true,
  linkcolor=blue,
  citecolor=blue,
  urlcolor=blue
]{hyperref}

\journal{Physics Letters B}

\begin{document}

\begin{frontmatter}



\title{Bayesian Inference for Extracting Barrier Distributions from Fusion Excitation Functions}


\author[FRIB,MSU]{Aaron Philip\corref{cor1}}
\ead{philipaa@mit.edu}
\cortext[cor1]{Corresponding author}

\author[FRIB]{Pablo Giuliani}
\ead{giulia27@msu.edu}

\author[FRIB]{Kyle Godbey}
\ead{godbeyky@msu.edu}

\affiliation[FRIB]{organization={Facility for Rare Isotope Beams},
            addressline={Michigan State University}, 
            city={East Lansing},
            postcode={48824}, 
            state={Michigan},
            country={USA}}
            
\affiliation[MSU]{organization={Department of Physics and Astronomy},
            addressline={Michigan State University}, 
            city={East Lansing},
            postcode={48824}, 
            state={Michigan},
            country={USA}}
            
\begin{abstract}
Barrier distributions encode rich information about the structure and dynamics of fusing nuclei, but extracting them from experimental fusion cross sections requires an estimate of the fusion excitation function's second derivative. In this work we approach the task of extracting barrier distributions with uncertainty estimates from sparse experimental measurements as a Bayesian inference problem. We introduce a method based on AutoBNN, an interpretable Bayesian machine learning framework, to provide a robust statistical approach for analyzing fusion excitation functions. Benchmarking against Gaussian process regression on simulated excitation functions that span a wide range of realistic experimental conditions, we find that AutoBNN more faithfully recovers the underlying barrier distribution and reports well-calibrated uncertainties. We then apply the AutoBNN method to four experimentally measured heavy-ion fusion reactions where it mitigates spurious above-barrier structure and constrains existing predictions. Alongside these results, we have developed a user-friendly software implementation of our method, facilitating its application to future heavy and light-ion fusion experiments.
\end{abstract}




\end{frontmatter}




\section{Introduction}
\begin{figure}[t]
    \centering
    \includegraphics[width=0.95\linewidth]{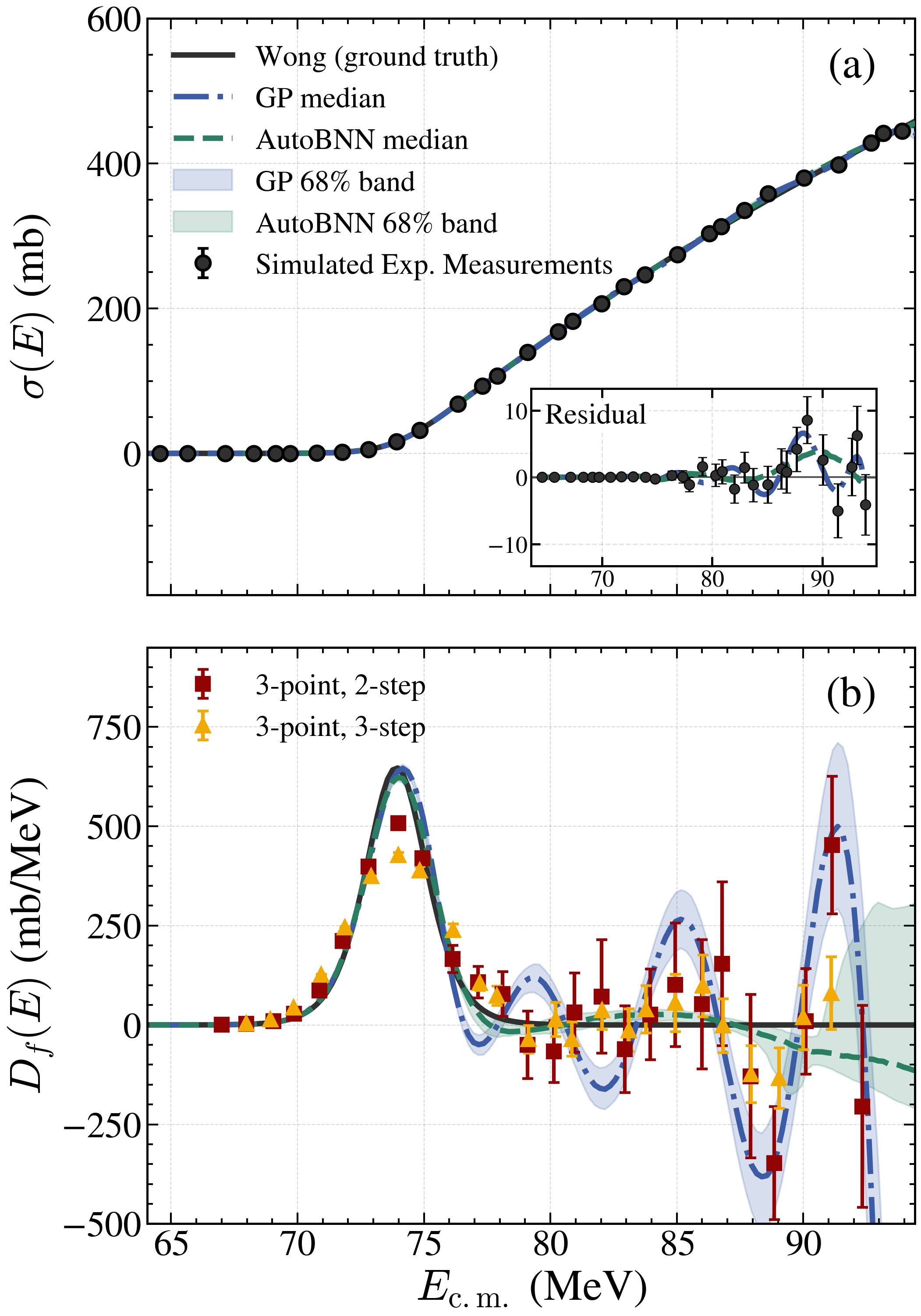}
    \caption{Comparison of several strategies to extract barrier distributions on a simulated excitation function. (a) AutoBNN and Gaussian process with RBF kernel fitted to simulated excitation functions generated using the Wong formula. (b) The barrier distribution extracted via the AutoBNN model, the GP, and the 3-point difference formula with two standard spacing choices of approximately 2 MeV (2-step) and 3 MeV (3-step), all compared to the ground truth Wong. 
    Error bars indicate $\pm 1\sigma$.}
    \label{fig:SampleFits}
\end{figure}

Studying the fusion of atomic nuclei across the nuclear chart is of fundamental scientific interest.
Heavy-ion fusion reactions are paramount to the production of rare isotopes and superheavy elements \cite{Back2014}, while light-ion fusion is relevant to  understanding stellar nucleosynthesis processes such as carbon burning \cite{Patterson1969} as well as the long-term pursuit of clean nuclear energy. 

The dynamics of fusion are influenced by the couplings between collective modes of the participating nuclei. The effect of these couplings can be interpreted as the presence of multiple tunneling barriers at different energies near the Coulomb barrier. Indeed, a single barrier model of fusion was found incapable of describing the sub-barrier fusion enhancement seen in heavy-ion fusion experiments in the 1980s \cite{Beckerman1988}. In Ref.~\cite{Rowley1991}, it was first demonstrated that this theoretical "distribution of barriers" could be linked to experimental fusion excitation functions $\sigma(E)$ via the (unnormalized) relation 
\begin{equation}
    D_f(E) = \frac{d^2}{dE^2}(E\sigma)
    \label{eq:bdist}
\end{equation} 

Barrier distributions reflect internal details of the fusing nuclei and their channel couplings like transfer or breakup as well as deformations, serving as information-rich signatures of nuclear structure \cite{Dasgupta1998}. Comparing barrier distributions from experimental data with those obtained from theoretical calculations provides a valuable testbed for theoretical models. For example, Coupled-channels calculations can investigate the couplings between transfer or breakup channels with entrance channels and collective excitations \cite{Balantekin1998}. Meanwhile, microscopic approaches like Time-Dependent Density Functional Theory can reflect the dynamics of deformations and intrinsic nucleonic densities as fusion takes place \cite{Umar2007} or even the effects of superfluidity arising due to pairing \cite{Scamps2018}. Recent experiments have revealed the importance of accounting for these channel couplings and collective effects \cite{Desouza2024, Desilets2025}, especially in the above-barrier region.

In practice, the second derivative in Eq.~\ref{eq:bdist} is estimated numerically from experimental measurements of the fusion excitation function $\sigma(E)$ sampled at different collision energies $\{E_i\}$. 
One common approach is to apply the established 3-point difference formula~\cite{Leigh1995, Dasgupta1998Review} to estimate the barrier distribution at discrete locations $E' = (E_{i-1} + 2E_{i} +E_{i+1})/4$
given by\footnote{Parts of the literature (see for example Refs.~\cite{Timmers1998,rowley1992sub}) have reported an inverted sign for the three point formula, likely as a typo.}
   
 \begin{equation}
   D_f(E') \approx 2[\frac{(E\sigma )_{i+1} - (E\sigma)_{i}}{E_{i+1} - E_{i}} - \frac{(E\sigma )_{i} - (E\sigma)_{i-1}}{E_{i}-E_{i-1}}](\frac{1}{E_{i+1} - E_{i-1}})
    \end{equation}
with an approximate uncertainty of \begin{equation}
    \delta D_f = (\frac{E'}{\Delta E^2})[(\delta \sigma)^2_{i-1}+4(\delta \sigma)^2_{i}+(\delta \sigma)^2_{i+1}]^{1/2}
\end{equation}
in the case that the spacing between data points is nearly constant i.e. $\Delta E = E_{i+1} - E_{i} = E_{i} - E_{i-1}$ \cite{Dasgupta1998Review}.
This approximation for $D_f(E')$ will be increasingly accurate if the spacing between measurements $\Delta E$ is smaller. However, the uncertainty in the prediction $\delta D_f$ also grows rapidly as $\Delta E$ becomes smaller.  Indeed, as noted in Ref.~\cite{Dasgupta1998}, larger step sizes will reduce the impact of an erroneous measurement but will also attenuate individual peaks in the barrier distribution which reduces the signal-to-noise ratio. Aside from this challenge, the 3-point formula can only estimate the barrier distribution at limited discrete locations and relies on differentiating measurements directly, amplifying the impact of measurement noise in the resulting barrier distribution. Given that the cross section spans orders of magnitude, the difference formula can be unreliable, particularly at energies above the Coulomb barrier. 

Other approaches involve fitting a model of choice to the cross section and differentiating the fitted excitation function. Recent works have sought to do so in a model-agnostic way to avoid encoding model-dependent assumptions in the resulting barrier distribution. Ref.~\cite{Scamps2018} introduced a local regression technique that fits piecewise polynomials around individual experimental data points to subsequently differentiate while Ref.~\cite{Jiang2022} approximated the barrier distribution as a sum of Gaussians fitted to measurements of the excitation function. Building on these ideas, Ref.~\cite{Godbey2024} introduced a method to recover a probability distribution of plausible barrier distributions by conditioning a Gaussian process on experimental measurements and quoted uncertainties of the excitation function. This approach provided an attractive alternative to the 3-point formula by replacing a discrete, noise-amplifying finite difference estimate with a continuous posterior capable of estimating uncertainties everywhere within the relevant energy range. However, Ref.~\cite{Philip2026} reported that, under realistic experimental conditions, this method can struggle to faithfully capture the barrier distribution and can introduce numerical aliasing of false peaks alongside inaccurate uncertainty estimates. 

More sophisticated statistical modeling approaches show promise in overcoming these challenges. In this work, we introduce a novel method for extracting barrier distributions with quantified uncertainties by utilizing AutoBNN \cite{AutoBNNSoftware2024}, an interpretable Bayesian machine learning tool originally developed for time series forecasting \cite{AutoBNN2024}. We benchmark the AutoBNN-based approach on simulated excitation functions under a host of realistic experimental conditions and find that AutoBNN models categorically outperform RBF-kernel Gaussian process models at recovering the barrier distribution alongside meaningful uncertainty estimates. Fig.~\ref{fig:SampleFits} illustrates these conclusions on a controlled benchmark problem: although each method may be fit to measurements of the known fusion excitation function $\sigma(E)$ with little visible variation, each method extracts substantially different barrier distributions $D_f(E)$. In particular, the 3-point difference formula and Gaussian process can struggle to faithfully recover the true barrier distribution everywhere within their quoted uncertainties, whereas the AutoBNN model excels in both the sub-barrier and above-barrier regions. 

The remainder of this Letter is organized as follows. In Section~\ref{sec: Methods}, we introduce our proposed Bayesian inference framework, Gaussian process regression, the AutoBNN tool, and the Wong formula which we used to benchmark our framework. In Section~\ref{sec:Results}, we present a comprehensive analysis of how both methods perform under a variety of realistic experimental conditions simulated using the Wong formula. We find that AutoBNN more accurately recovers the underlying statistical model in all scenarios. We then showcase barrier distributions extracted from heavy-ion fusion experiments using the AutoBNN framework. These results help interpret and constrain previously reported findings in the literature. We close by discussing the usage of this method in Section~\ref{sec:Conclusion}.

\section{Methods}\label{sec: Methods}
We present a generalization of the method introduced in Ref.~\cite{Godbey2024} that can incorporate other Bayesian inference techniques including the AutoBNN tool \cite{AutoBNN2024}.

\subsection{Bayesian Framework}
We formulate the problem of extracting barrier distributions from experimental data as a Bayesian inference task~\cite{Gelman2004d}. Namely, we treat the $n$ experimental measurements of the excitation function $\mathcal D_{\sigma}=(\mathbf{E}, \mathbf{y}, \mathbf{\delta y})$ with $\mathbf{E}, \mathbf{y}, \mathbf{\delta y} \in \mathbb{R}^n$ as noisy observations of some underlying smooth excitation function $\sigma(E)$ from which we hope to recover the barrier distribution via Eq.~\ref{eq:bdist}. 
Within our Bayesian framework we therefore seek to construct a posterior distribution given the experimental data: $ p(\sigma | \mathcal D_{\sigma})$. We can then sample cross section curves $\sigma^{(s)}(E)$ from which other quantities of interest, such as barrier distributions, can be obtained alongside uncertainty estimates.

In this work, we found it helpful to work with the logarithm of the data by determining the related posterior $p(g|\mathcal D)$ where $g(E)= \log(\sigma(E)/\bar y)$
with $\bar y=1$ mb. We define transformed dimensionless observations
\(z_i=\log (y_i/\bar y)\) with corresponding uncertainties approximated by the delta method \cite{Casella2024} as
\(\delta z_i \approx \delta y_i /y_i\). Thus the observational dataset we work with in practice is defined as $\mathcal D = (\mathbf{E}, \mathbf{z}, \mathbf{\delta z})$.

To construct our likelihood model, we assume an additive error model for the observations:
\begin{equation}
\begin{aligned}
    z_i &= g(E_i) + \epsilon_i, \\
    \boldsymbol{\epsilon} &\sim \mathcal{N}(\boldsymbol{0}, \Sigma_{\epsilon})
    \label{eq:additive error model}
\end{aligned}
\end{equation}
with $\Sigma_{\epsilon}$ diagonal, ensuring that observational noise $\boldsymbol{\epsilon}$ is independent among measurements. 
We determine the posterior $p(g | \mathcal D)$ with Bayes' theorem:
\begin{equation}
    p(g | \mathcal D) = \frac{p(\mathcal D| g)p(g)}{p(\mathcal D)}
\end{equation}

The \textit{prior} term $p(g)$ reflects initial knowledge about what functions $g$ may reasonably compose the distribution before any data is observed. The \textit{likelihood} term $p(\mathcal D| g)$ describes how well the data is explained by a function $g$ under the assumed statistical model given by Eq.~\ref{eq:additive error model}. Observational noise is assumed to be independent among measurements, so $p(\mathcal D| g)$ factors as 
\begin{equation}
    p(\mathcal D|g) = \prod_{i=1}^n p(z_i | g(E_i), \delta z_i ) .
    \label{eq:likelihood-factorized}
\end{equation}
The \textit{evidence} term $p(\mathcal D)$ is often interpreted as a normalization factor and is generally challenging to compute explicitly. 

After sampling functions $g^{(s)}(E)$ from the posterior $p(g | \mathcal D)$, these functions are transformed to obtain sampled barrier distributions as

\begin{equation}
\begin{aligned}
    \sigma^{(s)}(E) = \bar y\exp[g^{(s)}(E)], \\
    D^{(s)}_f(E) = \frac{d^2(E\sigma^{(s)})}{dE^2}
\end{aligned}
    \label{eq:barrier-sampling-extraction}
\end{equation}

A mean or median prediction for the barrier distribution with uncertainties can be formed by repeatedly sampling functions $D_f^{(s)}(E)$. We note that working with an additive error model in logarithmic space is equivalent to assuming a multiplicative error model of cross section measurements in linear space. Multiplicative error models are often used in contexts where observations are always positive and uncertainties scale with the observations and span many orders of magnitude \cite{Limpert2001} as is true for fusion cross section data.

We now introduce Gaussian processes and AutoBNN, then detail how we sample the posterior $p(g|\mathcal D)$ using each method. 

\subsection{Gaussian process regression}
Gaussian process regression has become a ubiquitous tool in machine learning and data-driven modeling. Formally, a Gaussian process (GP) is defined as a set of random variables of which the distribution of any finite subset forms a multivariate Gaussian \cite{RasmussenWilliams2006}. After conditioning on data, sampling from a GP is commonly interpreted as sampling functions $g$ from a posterior $p(g | \mathcal D)$.

In the one-dimensional case, a conditioned GP can sample functions from $p(g | \mathcal D)$ by discretizing $g(E)$ into a set of $N$ values $\mathbf{g}_{*} \in \mathbb{R}^N$ defined on a (typically high-resolution) grid $\mathbf{E}_{*}\in \mathbb{R}^N$ and determining a multivariate Gaussian distribution over the elements of $\mathbf{g}_{*}$. This distribution is conditioned on the $n$ observations in $\mathcal D$. By assuming additive Gaussian noise as in Eq.~\ref{eq:additive error model} and defining $\Sigma_{\epsilon} = \operatorname{diag}(\delta z_1^2,\ldots,\delta z_n^2)$, the likelihood is also normally distributed and given by Eq.~\ref{eq:likelihood-factorized}. Therefore, the posterior over the discretized function is available in closed form \cite{Goldberg1997} as the multivariate distribution: 
\begin{equation}
\begin{aligned}
p(\mathbf{g}_{*} \mid \mathcal D) = \mathcal N([&\,\mathbf{m}_{*} + K(\mathbf{E}_*,\mathbf{E})K_z^{-1}(\mathbf{z} - \mathbf{m})], \\
&\,[K(\mathbf{E}_*,\mathbf{E}_*) - K(\mathbf{E}_*, \mathbf{E})K_z^{-1}K(\mathbf{E},\mathbf{E}_*)])
\end{aligned}
\label{eq: GP}
\end{equation} 
where $K_z = K(\mathbf{E},\mathbf{E}) + \Sigma_{\epsilon}$ and $\mathbf{m} \in \mathbb{R}^n$,  $\mathbf{m}_{*}\in \mathbb{R}^N$ are evaluations of the prior mean function $m(x)$ at the observed and inference locations respectively. In this work, we use the zero-mean prior $m(x)=0$. An essential prior also arises through the choice of the kernel function $k(x, x')$ that reflects the modeler's beliefs about the correlation structure between nearby function values. This kernel specifies the covariance matrices $K(\mathbf{*},\mathbf{*})$: for example, $K(\mathbf{E},\mathbf{E})_{i,j} = k(E_{i}, E_j)$. 
We utilize a Radial Basis Function (RBF) kernel for all experiments in this paper. For smooth kernels such as the RBF used here, posterior function samples can be evaluated on a dense grid and numerically transformed according to Eq.~\ref{eq:barrier-sampling-extraction}. 
 
The use of GPs is attractive because they are easy to obtain posterior samples from, they provide uncertainty estimates on the extracted barrier distribution, and they directly incorporate quoted experimental uncertainties. However, standard smooth kernels are often prone to introducing numerical artifacts in the barrier distribution, accompanied by overconfident and unreliable uncertainty estimates \cite{Philip2026}.

\subsection{AutoBNN}
Bayesian Neural Networks (BNNs) operate similarly to standard neural networks, except their internal parameters are sampled from distributions. If the parameter distributions are built within a robust statistical framework that accurately models the sources of noise and uncertainty in the data, the BNN predictions will provide meaningful uncertainty quantification. Ref.~\cite{Neal1996} showed that BNNs converge to Gaussian processes in their infinite width limit. More recent research \cite{Meronen2021} has explicitly determined which activation functions and weight initializations for wide BNNs allow them to mimic the correlation structures of several standard GP kernels. Different BNN surrogates for common GP kernels can then be combined with operations analogous to adding, multiplying, or concatenating GP kernels \cite{pearce2019}.

The AutoBNN package \cite{AutoBNNSoftware2024} was first introduced by Google in the context of time series forecasting \cite{AutoBNN2024} and unified many of these advances in connecting BNNs to Gaussian processes \cite{Meronen2021, pearce2019, Saad2023}. AutoBNN proposes to automate model discovery by combining BNN surrogates for many common GP kernels to support richer function priors. The AutoBNN package presents several predefined weighted combinations of these BNN surrogates which can explore a combinatorially large space of statistical models while training. 

We leverage these model-discovery architectures to determine a posterior distribution of functions $p(g | \mathcal D)$. Before training, AutoBNN places priors over parameters $p(\theta)$ of the predefined architecture $g_\theta(E) = \mathcal C[\psi_{\theta}^{(1)}(E), ..., \psi_{\theta}^{(M)}(E)]$ where $\mathcal C$ is some combination operator and $\psi_{\theta}^{(i)}(E)$ denotes individual BNN surrogates for GPs with common kernels $k^{(i)}(E,E')$. The parameters are then directly conditioned on observations $\mathcal D$ through Bayes' rule for the parameters: 

\begin{equation}
   p(\theta | \mathcal D) =   \frac{p(\mathcal D | \theta)p(\theta)}{p(\mathcal D)}
\end{equation}

The parameters are sampled from the posterior $\theta^{(s)} \sim p(\theta | \mathcal D)$ which induces a distribution over corresponding functions $ g_{\theta}^{(s)}(E)$. These smooth latent functions are transformed via Eq.~\ref{eq:barrier-sampling-extraction} to obtain samples $D^{{(s)}}_f(E)$. A key difference from the GP method is that because the AutoBNN models are implemented in JAX, we can exploit forward-mode automatic differentiation to compute each $D^{(s)}_f(E)$ without finite difference errors.

We sample from the posterior $p(\theta | \mathcal D)$ through Markov Chain Monte Carlo (MCMC) using the No-U-Turn Sampler algorithm \cite{Hoffman2014}, a modification of Hamiltonian Monte Carlo \cite{Duane1987} which integrates equations of motion for each parameter in the network. For all our experiments, we utilized the predefined Sum of Changepoint architecture in the AutoBNN package~\cite{AutoBNNSoftware2024}. The Sum of Changepoint models concatenate piecewise kernel combinations together, which naturally enables the BNNs to reflect different statistical structures in different energy ranges. Consequently, these models are occasionally prone to unphysical discontinuities in the barrier distribution. We largely suppressed these effects by penalizing large values of the third derivative of each $g_{\theta}^{(s)}(E)$ and therefore each $\sigma^{(s)}(E)$ with a modified prior we introduced into the framework:

\begin{equation}
\begin{aligned}
    p(\theta) &\to
    p(\theta)
    \exp(-\lambda L[g_{\theta}]),
    \\
    L[g_{\theta}]
    &=
   \frac{1}{E_{max}-E_{min}} \int_{E_{min}}^{E_{max}}
    \left[
    \frac{d^3 g_{\theta}}{dE^3}
    \right]^2
    dE .
\end{aligned}
\end{equation}

In practice, we compute $L[g_\theta]$ numerically by evaluating $g_{\theta}$ on a high-resolution grid. The modified prior discourages function samples with discontinuities in the barrier distribution over the relevant energy range $[E_{min},E_{max}]$. The strength of the prior can be tuned with $\lambda$ which we set to 1.0 $\mathrm{MeV}^6$ for all experiments. 

\begin{figure*}[t]
    \centering
    \includegraphics[width=\linewidth]{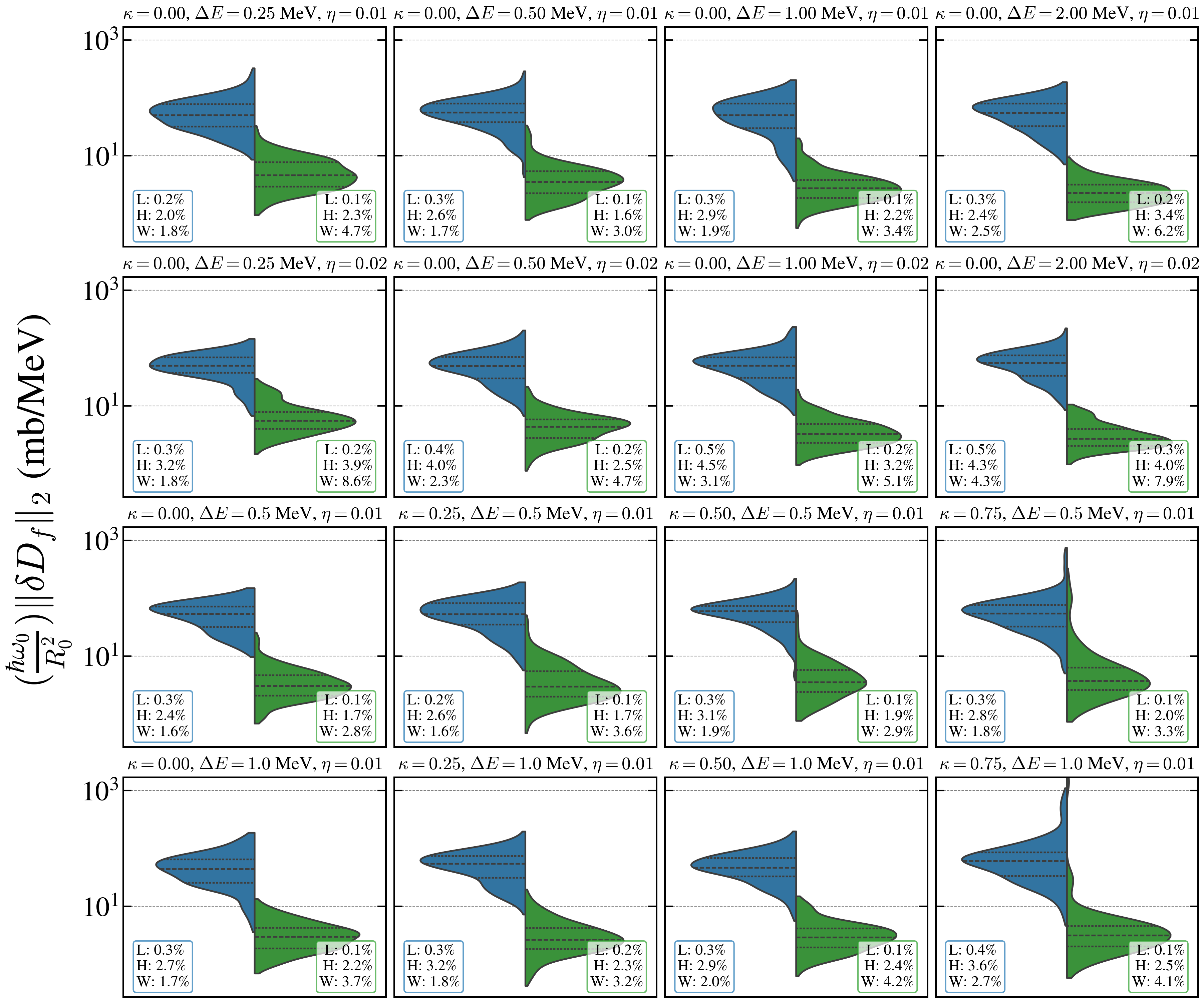}
    \caption{Distributions of normalized error residuals of AutoBNN (green) and Gaussian process (blue) models in recovering the underlying ground truth Wong barrier distribution across 16 different simulated experimental conditions. The insets provide more local metrics on how well both models recovered the location (L), height (H), and width (W) of the primary barrier. See Section~\ref{subsec:dataset-construction} for additional details about each dataset's properties.}
    \label{fig:MAEGrid}
\end{figure*}
\begin{figure}[t]
    \centering
    \includegraphics[width=\linewidth]{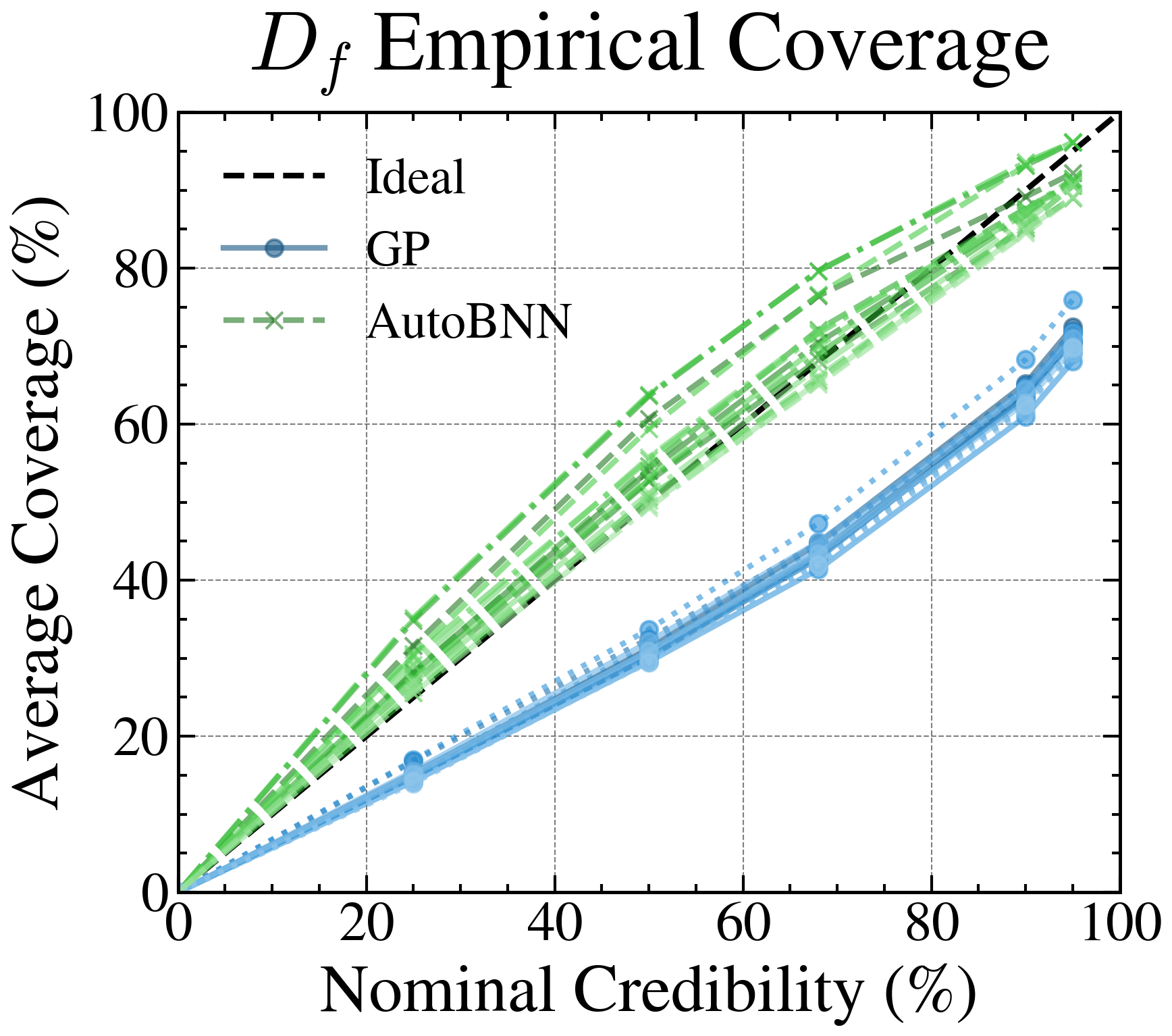}
    \caption{The average empirical coverage of AutoBNN models and Gaussian process models across 16 simulated datasets. Each green (blue) line corresponds to a single dataset that 100 AutoBNN models (Gaussian process models) were trained to before being evaluated at several nominal credible intervals. Optimal coverage corresponds to a line lying on the diagonal.
    }
    \label{fig:Coverage}
\end{figure}

\subsection{Wong Formula}
To benchmark the two Bayesian inference methods, we generated realistic data based on the simple model of fusion given by the Wong formula \cite{Wong1973}: 

\begin{equation}
    \sigma(E) = \frac{R_0^2\hbar\omega_0}{2E} \log(1+
    \exp[2\pi (E-E_0)/\hbar \omega_0]),\label{eqn:wong}
\end{equation} 
which is parameterized by the radius $R_0$, height $E_0$, and curvature $\hbar \omega_0$ of the approximately parabolic barrier in the s-wave for a Woods-Saxon-like interaction. 

The corresponding barrier distribution for this model is then given in closed form by
\begin{equation}
    D_f(E) = \frac{2\pi^2R_0^2}{\hbar \omega_0 }\frac{e^x}{(1+e^x)^2}
\end{equation}
with $x = \frac{2\pi}{\hbar \omega_0}(E-E_0)$ \cite{Rowley1991}.

\begin{figure*}[t]
    \centering
    \includegraphics[width=\linewidth]{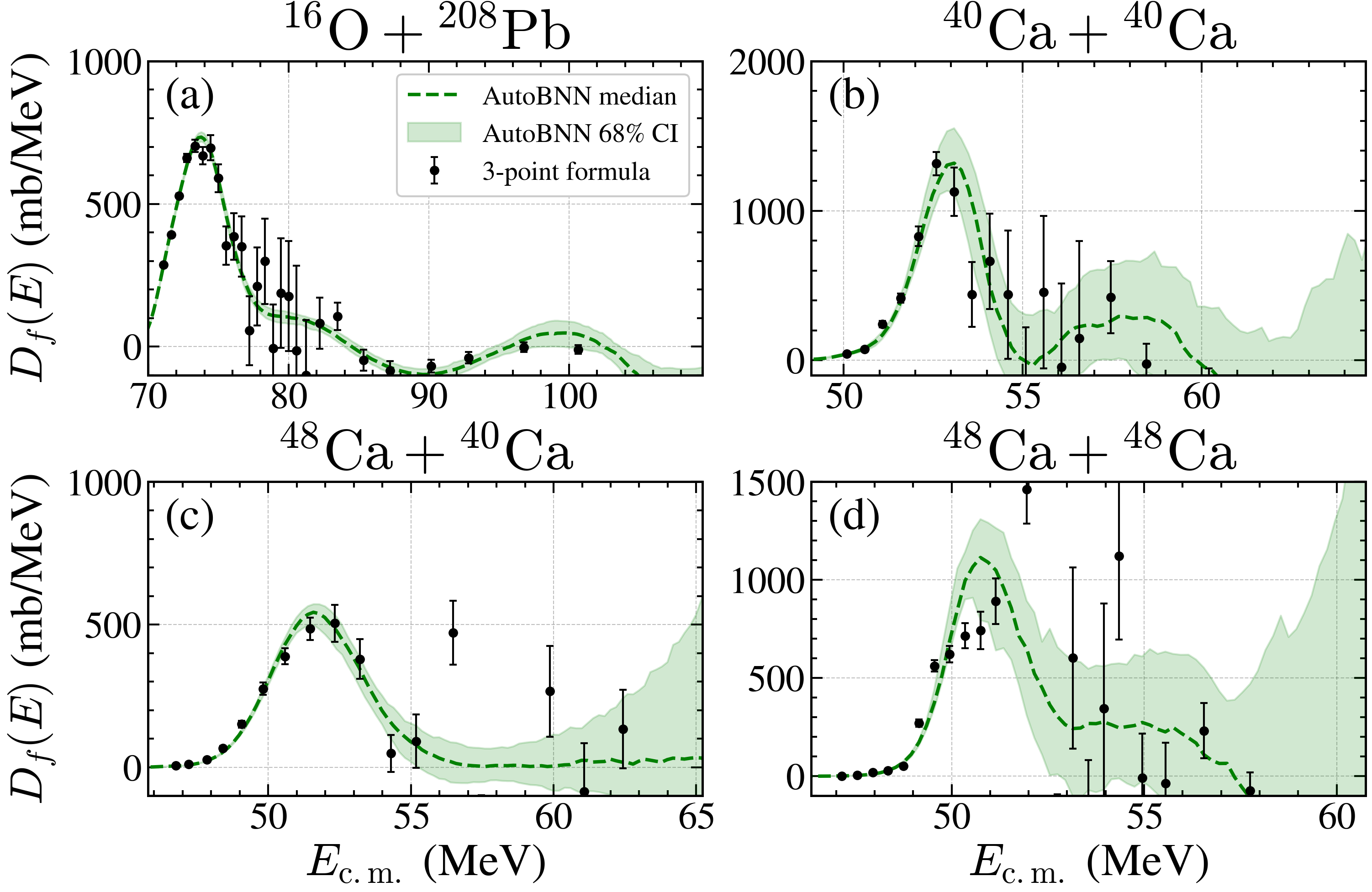}
    \caption{Barrier distribution extractions from real fusion experiments reported in (a) Morton et al. \cite{Morton1999}, (b) Montagnoli et al. \cite{Montagnoli2012}, (c) Jiang et al. \cite{Jiang2010}, and (d) Stefanini et al. \cite{Stefanini2009} using the AutoBNN framework alongside discrete predictions made using the 3-point formula with a 2-step difference and error bars estimating 1 standard deviation.}
    \label{fig:ExperimentalFits}
\end{figure*}

\subsection{Dataset Construction}
\label{subsec:dataset-construction}
To test the relative merits of the GP and AutoBNN-based methods for extracting barrier distributions from experimental data, we built an analysis pipeline modeled after that in Ref.~\cite{Yan2018}, in which the authors studied the performance of various functions to extract the proton radius from scattering data. By varying the irregularity in spacing, resolution, and noise within realistic bounds, we investigated the relative performance of the GP and AutoBNN techniques in extracting the underlying ground truth barrier distribution. We built 16 datasets of 100 curves per dataset characterized by three quantities: 
\begin{enumerate}
    \item $\kappa$: The coefficient of variation in the energy spacing of $E_\mathrm{c.m.}$. Higher values of $\kappa$ correspond to increasingly irregular measurements of the excitation function. 
    
    \item $\Delta E$: The average spacing of $E_\mathrm{c.m.}$ in MeV. In the case where $\kappa=0$, the excitation function is sampled uniformly at a resolution of $\Delta E$.
    
    \item $\eta$: The percent error of each measurement $\sigma(E_i)$ on the excitation function.
\end{enumerate}
Based on an analysis of several heavy-ion fusion experiments \cite{Morton1999, Montagnoli2012, Jiang2010, Stefanini2009, Dasgupta2007, Aljuwair1984, Trotta2001}, we chose realistic ranges for these quantities. Each curve is generated from a distinct random parameter sample of $(E_0, R_0, \hbar\omega_0)$ according to the values of $(\kappa, \Delta E, \eta)$. Each Wong formula parameter was sampled uniformly in the ranges $E_0 \in [50.0, 80.0]$ MeV, $R_0 \in [8.0, 15.0]$ fm, $\hbar \omega_0 \in [2.0, 8.0]$ MeV. As such, each dataset contained a diverse representation of many different barrier shapes and locations. Given the computational demands of training AutoBNN models and the strong summary statistics we obtained, we limited ourselves to 100 curves per dataset. 

For each Bayesian inference method, we conditioned a distinct model to each curve in a dataset. We utilized GPs with Radial Basis Function kernels and the predefined Sum of Changepoint models in the AutoBNN package. The limited amount of observational data per curve ensured that each GP only took on the order of minutes to train. We trained each AutoBNN model with 4 Markov chains with 1000 burn-in steps followed by 1000 sampling steps which typically required 3-4 hours per model on a single CPU.

\section{Results and Discussion}
\label{sec:Results}
\subsection{Benchmarking Study}
Fig.~\ref{fig:MAEGrid} summarizes the performance of the median barrier distribution of each GP and AutoBNN fit. For each curve, we took the L2 norm of the residual of the barrier distribution on the entire energy range provided and normalized by the prefactor of the true barrier distribution to enable comparisons across different Wong parameter samples. Each of the insets also provides more local metrics on how well each model recovered the location (L), height (H), and width (W) of the primary Wong peak on average. A typical example of the true Wong distribution can be seen in Fig.~\ref{fig:SampleFits}b. 

In all 16 cases, the residual distributions indicate that the AutoBNN models surpass the GP models' performance in capturing the true barrier distribution over the full energy interval. Notably, AutoBNN models outperform GPs more dramatically on the more challenging datasets with greater irregularity, lower resolution, and/or higher amounts of noise. Interestingly, the GPs often recover the height and width of the primary peak more accurately while AutoBNN models excel at capturing the peak location. 

The empirical coverage plot in Fig.~\ref{fig:Coverage} clearly establishes that the AutoBNN models also outperform the GP models in estimating their corresponding uncertainties. Each line in Fig.~\ref{fig:Coverage} corresponds to one of the 16 datasets, where each dataset is characterized by $(\kappa, \Delta E, \eta)$ and contains 100 curves generated with varied parameters of the Wong formula. The coverage for a single curve is defined as the percentage of the true Wong formula curve that lies within the nominal credibile band over the entire interval. The 25\%, 50\%, 68\%, 90\%, and 95\% coverages were averaged over all curves in each dataset to plot each line in Fig.~\ref{fig:Coverage}. Overall, the BNNs are much closer to the diagonal with a slight bias towards conservatively underconfident predictions. In sharp contrast, the GPs all lie well below the diagonal indicating a poor, overconfident estimate of their uncertainties. As supported by Fig.~\ref{fig:MAEGrid}, part of this reason is because the mean of the GPs is typically farther from the true barrier distribution than the AutoBNN models. Notably, the GPs perform very consistently across all datasets.

Taken together, Figs.~\ref{fig:MAEGrid} and ~\ref{fig:Coverage} establish that across a wide range of realistic scenarios, the AutoBNN Sum of Changepoint models robustly extract the underlying barrier distribution while quoting well-calibrated uncertainty estimates. 

\subsection{Heavy-ion Fusion Reaction Analysis}
Having thoroughly tested the performance of AutoBNN on a toy problem, we now apply the method to well-studied fusion experiments. Fig.~\ref{fig:ExperimentalFits} displays the barrier distributions extracted from four well-known fusion experiments \cite{Morton1999, Montagnoli2012, Jiang2010, Stefanini2009} that were also studied in Ref.~\cite{Godbey2024} to probe the effect of different amounts of nucleonic transfer in closed-shell nuclei. We include predictions from the 3-point difference formula with a 2-step difference for the sake of comparison. The AutoBNN results are especially useful to both support and help constrain the extracted barrier distributions in Ref.~\cite{Godbey2024} as well as the 3-point estimates reported in the original experimental papers. 

We selected all predictions by fitting a Sum of Changepoint model to a single experimental dataset multiple times and selecting the fit with the highest mean value of the posterior over all sampled functions. Identifying a more robust selection metric is an open question, but our preliminary search through established estimators and scoring rules \cite{gneiting2007} was inconclusive.

In our analysis of the data presented in Ref.~\cite{Morton1999} on $^{16}$O $+^{208}$Pb (panel (a) of Fig.~\ref{fig:ExperimentalFits}), the AutoBNN and 3-point methods are closely aligned across the entire energy range. The interpolation provided by the AutoBNN method offers greater insight into the above-barrier salient structure emerging around 80 MeV and 100 MeV while constraining the 3-point formula's reported uncertainty around 80 MeV.

The AutoBNN median prediction in the $^{40}\mathrm{Ca} + ^{40}\mathrm{Ca}$ reaction in Ref.~\cite{Montagnoli2012}  (panel (b) of Fig.~\ref{fig:ExperimentalFits}) supports the presence of a secondary bump around 58 MeV as also indicated by both Coupled-channels calculations and the 3-point formula in Fig. 2 of Ref.~\cite{Montagnoli2012}. Near the primary peak at 53 MeV, the AutoBNN partially resolves the structure and location of the peak while indicating that the peak is not rigidly constrained by the data. At larger energies, the AutoBNN uncertainty grows but the model offers a more complete picture of the barrier distribution around 55 MeV where the 3-point formula alone is uninformative.

In analyzing the $^{48}\mathrm{Ca} + ^{40}\mathrm{Ca}$ reaction reported in Ref.~\cite{Jiang2010}   (panel (c) of Fig.~\ref{fig:ExperimentalFits}) and the $^{48}\mathrm{Ca} + ^{48}\mathrm{Ca}$ reaction in Ref.~\cite{Stefanini2009}   (panel (d) of Fig.~\ref{fig:ExperimentalFits}), the AutoBNN and 3-point formula predictions more obviously disagree. We find that the measurements in Ref.~\cite{Jiang2010} support the presence of a single smooth peak per the AutoBNN extracted barrier distribution. In contrast, the 3-point formula indicates a peak around 56 MeV of approximately equal magnitude to the primary peak. Interestingly, the GP prediction in Ref.~\cite{Godbey2024} is also more closely aligned with the AutoBNN prediction. 

For the $^{48}\mathrm{Ca} + ^{48}\mathrm{Ca}$ reaction in Ref.~\cite{Stefanini2009}, the fitted AutoBNN model predicts a substantially different structure and location for the primary peak compared to the 3-point formula. At higher energies, the distribution is poorly constrained by both models. It is challenging to reconcile the 3-point predictions above the Coulomb barrier with the AutoBNN predictions. The 3-point formula predicts a sharp secondary rise in the barrier distribution around 55 MeV while the AutoBNN supports a more gradual secondary plateau near 55 MeV. 

Figure~\ref{fig:SampleFits}b illustrates an example in which the 3-point difference formula often struggles to faithfully recover the barrier distribution with meaningful uncertainties compared to the AutoBNN method. Since this behavior appears ubiquitously across the various scenarios we explored (see Ref.~\cite{Philip2026} for more details), we therefore place a greater emphasis on predictions made by the AutoBNN models. 

\section{Conclusion}
\label{sec:Conclusion}

In this Letter, we introduced and benchmarked a method using the AutoBNN Bayesian machine learning framework for extracting barrier distributions from fusion excitation functions.
This task is challenging because experimentally measured cross sections are discrete, span many orders of magnitude, and carry heteroscedastic uncertainties, while the barrier distribution depends on a potentially numerically unstable second derivative of the energy-weighted cross section $E\sigma(E)$.

Using realistic simulated data generated with the Wong formula, we compared the performance of the new AutoBNN method with a previous approach based on Gaussian process regression introduced in Ref.~\cite{Godbey2024}. The Gaussian process models with a standard RBF kernel regularly introduce spurious oscillations with poorly calibrated uncertainties in the extracted barrier distribution.
By contrast, the Sum of Changepoint architecture in the AutoBNN approach consistently recovered the true barrier distribution more faithfully and quoted well-calibrated uncertainties across all datasets.
We then applied the AutoBNN framework to four well-studied heavy-ion fusion reactions.
The extracted barrier distributions often agree with the three-point difference formula below and near the primary peak, diverge from it above the barrier as anticipated, and avoid the spurious above-barrier structure to which both the three-point and GP methods are prone. 

We anticipate that the method we developed in this work will be valuable for the re-analysis of previous data, as well as for upcoming heavy- and light-ion fusion experiments. As such, we have included a guided example on how to apply this tool to new experiments in the linked GitHub repository \cite{PhilipRepo2026}. One of the potential benefits of the continuous posterior offered by AutoBNN is the opportunity to precisely gauge where future experimental data may be maximally impactful. An improved ability to extract reliable barrier distributions with quantified uncertainties will help the nuclear science community capitalize on the discovery potential of new fusion reaction data.

\section*{Declaration of competing interest} The authors declare that they have no known competing financial interests or personal relationships that could have appeared to influence the work reported in this paper.

\section*{Data availability} The data and code used in this work are available at \url{https://github.com/aaron-philip/barrier-extraction}.

\section*{Declaration of generative AI and AI-assisted technologies in the manuscript preparation process}
During the preparation of this work, the authors utilized OpenAI's Codex service to optimize, generate, and review code. After using this tool, the authors reviewed and edited all content and take full responsibility for the content of the published article. No written content was generated using any form of AI, but typographical errors and spelling were detected with LLMs.

\section*{Acknowledgements}
This work was supported by the Department of Energy (DOE) under grant numbers DOE-DE-SC0026198 (STREAMLINE 2 Collaboration) and DOE-DE-NA0004074 (NNSA, the Stewardship Science Academic Alliances program).


\bibliographystyle{elsarticle-num} 
\bibliography{BarrierPaperBib}






\end{document}